\documentclass[12pt, a4paper, times]{article}
\usepackage[english]{babel}
\usepackage[T1]{fontenc}
\usepackage{a4wide}
\usepackage{color}
\usepackage[centertags]{amsmath}
\usepackage{amsfonts}
\usepackage{amssymb}
\usepackage{amsthm}
\usepackage{newlfont}
\usepackage{amssymb,amsbsy,times,fancyhdr,color}
\usepackage{amsmath}
\usepackage[title]{appendix}
\usepackage{booktabs}
\usepackage{multirow}
\usepackage{subfigure}
\usepackage{breqn}
\usepackage{graphicx}
\usepackage{subfigure}
\usepackage{float}
\usepackage{multirow}
\usepackage{array}
\usepackage{threeparttable}

\newcommand{\tabincell}[2]{\begin{tabular}{@{}#1@{}}#2\end{tabular}}

\newcounter{licznik}[section]

\newtheorem{Theorem}[licznik]{THEOREM}

\newtheorem{Proposition}[licznik]{PROPOSITION}
\newtheorem{Algorithm}[licznik]{ALGORITHM}

\begin{document}
\title{Dynamic Dependence Modeling in financial time series}

\author{Yali Dou\textsuperscript{a},  H. Liu\textsuperscript{a} and G. Aivaliotis\textsuperscript{a,b}\\
\textsuperscript{a} School of Mathematics, University of Leeds, UK, ~~ \textsuperscript{b}\footnote{Contact: G.Aivaliotis@leeds.ac.uk} The Alan Turing Institute, UK
}

\maketitle

\maketitle

\begin{abstract}
This paper explores the dependence modeling of financial assets in a dynamic way and its critical role in measuring risk. Two new methods, called Accelerated Moving Window method and Bottom-up method are proposed to detect the change of copula. The performance of these two methods together with Binary Segmentation \cite{vostrikova1981detection} and Moving Window method \cite{guegan2009forecasting} is compared based on simulated data. The best-performing method is applied to Standard \& Poor 500 and Nasdaq indices. Value-at-Risk and Expected Shortfall are computed from the dynamic and the static model respectively to illustrate the effectiveness of the best method as well as the importance of dynamic dependence modeling through backtesting.

\end{abstract}

\section{Introduction}\label{sec:intro}
The dependence structure of financial assets has a significant effect on risk measurement thus its modeling is critical. While linear correlation is adequate in measuring dependency between financial assets under the assumption of elliptically distributed returns, it is not suitable when these are asymmetrical and heavy tailed, see Embrechts et al \cite{embrechts2002correlation}. Patton \cite{patton2004out} illustrated that the distribution of individual assets is asymmetric and the financial returns have stronger dependence during economic recessions than booms.
Embrechts et al. \cite{embrechts2001} introduced copulas to modeling dependence of financial data. A copula function, (see Sklar \cite{sklar1959fonctions}) joins the marginal distributions of individual assets into the joint distribution and encapsulates the dependence structure. Examples of copula functions used in finance are Gaussian copula, Student-t copula (Du and Lai \cite{du2017copula}), Clayton and Gumbel copula as well as their variants. However, the assumption that the dependence of financial assets remains constant is often unrealistic as experience has proven (see for example Li's model \cite{Li2000} in relation to the use of Gaussian copula and the financial crisis of 2008). For this reason, there is an incentive to model the dependency of the financial portfolio returns dynamically. The changes in the dependence structure can be either in the form of copula family change, parameter change or both at certain points in time and are usually related to changes in the market conditions. 


The use of static copula models is well established. H\"{u}rlimann \cite{hurlimann2004fitting}, proposed a copula-based statistical method and fitted three copula models that have best overall fit between Swiss Market Index and one of the stocks in the index successfully, while de Melo Mendes and de Souza \cite{de2004measuring} captured the dependence between the Brazilian and American markets by fitting a copula and demonstrated that the selection of the copula family has a significant impact on the results of risk measures such as Value at Risk (VaR) and Expected Shortfall (ES). Du and Lai \cite{du2017copula} verified that the dependence of offshore Renminbi (CNH), and onshore Renminbi (CNY) is asymmetrical, which strengthens the argument against linear and symmetric measures like linear correlation. Cheng et al. \cite{cheng2007new} modeled the dependency between Shanghai Stock Composite Index and Shenzhen Stock Composite Index with traditional Monte Carlo, pure copula method and mixture copula method. They concluded that the role of dependence structure in measuring risk is much more valuable than the marginal distribution of asset by comparing six risk measures including VaR and CVaR.  

Models that allow for parameter changes in the copula under the assumption of invariant copula family are more realistic than the static models and have been employed extensively. Patton \cite{patton2001modelling} introduced the concept of conditional copula based on an extension of Sklar's theorem \cite{sklar1959fonctions} and verified the time-varying dependence structure of exchange rates returns for Deutsche Mark quoted against U.S. dollar (DEM/USD) and Yen quoted against U.S. dollar (JPY/USD). Furthermore, evidence of existence of change points in the conditional copula of DEM/USD and JPY/USD was provided by Embrechts and Dias \cite{da2004change}, who also mentioned that the change points in copula were closely associated with financial events.
Bartram et al. \cite{bartram2007euro} demonstrated that the introduction of Euro (EUR) changed the dependency of European markets, especially those in France, Germany, Italy, the Netherlands and Spain. Jondeau and Rockinger \cite{jondeau2006copula} captured the parameter change of the copula between each pair of four stock indexes (S\&P 500, FTSE, DAX, and CAC) increasing the dependence between them. Dias and Embrechts \cite{dias2010modeling} fitted two copula models, time-varying and time-invariant respectively, for foreign exchange EUR/USD and JPY/USD and found that the time-invariant copula performed poorly in describing the tail behavior.

Caillault and Guegan \cite{guegan2004forecasting} first established the variation of copula family in measuring risk using a moving window approach. Zhang and Guegan \cite{zhang2006change} also applied the moving window approach and detected the change both in copula family and parameter of Asian foreign exchange markets, USD/THB (Thai Baht) and USD/MYR (Malaysian Dollar).  Guegan and Zhang \cite{guegan2010change} fitted a dynamic copula model at historical data using the Binary Segmentation \cite{vostrikova1981detection} method to determine the change time of copula family and parameter. They used a kernel-based goodness-of-fit test to detect the change of family while a change point analysis introduced by Embrechts and Dias \cite{da2004change} was utilized to find the change time of copula parameter. 

In this paper, we adopt a new rank-based goodness-of-fit test proposed by Huang and Prokhorov \cite{huang2014goodness}, which can detect both family and parameter changes of the copula. This test comes from the White \cite{white1982maximum} specification test and relates to the information matrix equality. Unlike White \cite{white1982maximum}, it is based on the empirical marginal distributions instead of parametric distributions. Compared with other ``blanket'' tests \cite{genest2009goodness} known, there is no need for any parameter adjustment or strategic choices. This goodness-of-fit test has an asymptotic distribution ($\chi^2$ distribution), which makes it much simpler in computation although it has the rather strict assumption of three-times differentiability of the log-copula function.


The most popular two approaches to detect the change of copula are the Binary Segmentation \cite{vostrikova1981detection} and the Moving Window method \cite{guegan2009forecasting}. The main idea of Binary Segmentation is splitting the whole sample into two subsamples if the copula does not remain the same repeatedly until this is not the case. The ``top-down'' feature of the Binary Segmentation made it easy to implement and compute efficiently. However, the drawback of this method is that the performance is poor in more complicated situations, see Fryzlewicz \cite{fryzlewicz2016tail}. On the contrary, Moving Window method fixes an initial window which is then moved forward until the whole sample is covered. This method can detect the change more precisely while the width of the moving window is still an open question, see Caillault and Guegan \cite{guegan2009forecasting}. Note that an essential difference between the two approaches is that binary segmentation is backward looking while the moving window can be used as "a real time" tool.

\begin{figure}[H]
\centering
\subfigure[]{
\label{Fig.sub.1}
\includegraphics[width=6cm]{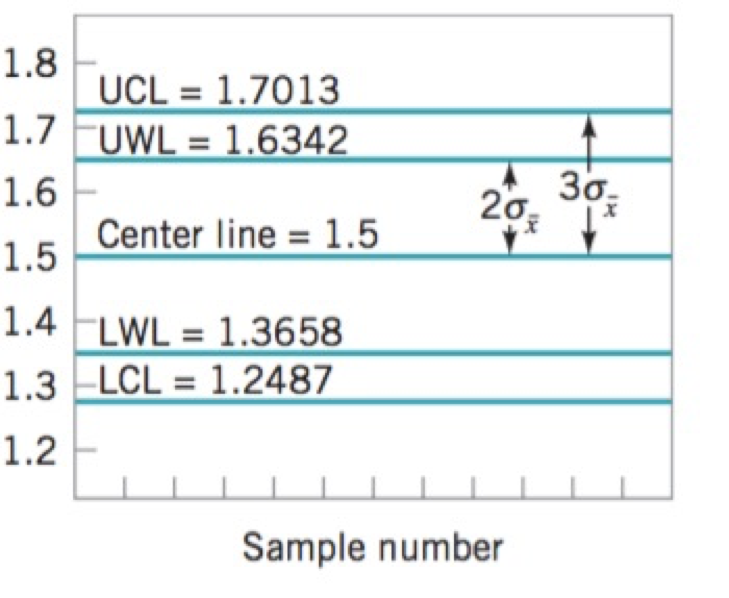}}
\subfigure[]{
\label{Fig.sub.2}
\includegraphics[width=6cm]{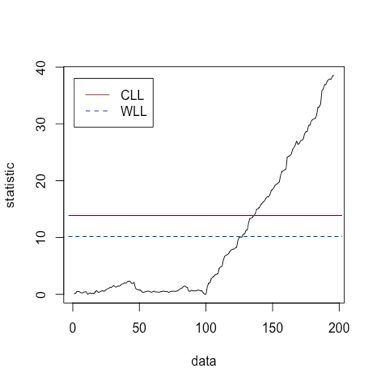}}
\caption{Picture (a) is the control chart that proposed by Montgomery \cite{montgomery2009statistical}. Picture (b) is the movement of the statistics that produced by the goodness-of-fit test.}
\label{Fig.intro}
\end{figure}

In this paper, we propose two new methods that we call Accelerated Moving Window and Bottom-up approach. The idea of Accelerated
 Moving Window came from the area of statistical quality control (see chart in Figure \ref{Fig.sub.1}, suggested by Montgomery \cite{montgomery2009statistical}) and the monotonic increase of the test statistic of the Huang and Prokhorov \cite{huang2014goodness} test when data from a copula different than the one of the null hypothesis are added. 
Combining Moving Window and control limit lines, the Accelerated Moving Window approach monitors the movement of the test statistic using the control limit line (CLL) as well as the warning limit line (WLL) (see Figure \ref{Fig.sub.2}). CLL and WLL are determined by the critical values of the  $\chi^2$ distribution at the confidence level $\alpha_w$ and $\alpha_c$ respectively, where $\alpha_w < \alpha_c$. 
Once the statistic exceeds the WLL, there is a signal that the copula may have changed. Then the length of the window shrinks and the test is repeated to see if CLL is breached. Once a change point is detected (CLL crossed), a new copula is fitted from this point (using a number of past data) and the process is iterated until the whole data is covered. 

In the field of backward looking methods, we propose an approach that we call "Bottom-up", which is inspired by the tail-greedy bottom-up method, see Fryzlewicz \cite{fryzlewicz2016tail}. The main idea is to divide the sample data into multiple small (how small to be determined by the power of the test) intervals with no change point in each interval, then merging the contiguous intervals layer by layer if they belong to the same copula until all with the same copula family have been merged. 

We compare the performance of these two new methods with Binary Segmentation and Moving Window in detecting the change in copula family and parameter. First, we simulate the data from two different bivariate copula families with the same parameter and evaluate the behavior of these four methods in detecting the copula family change. Contrary to Binary Segmentation and Bottom-up approach, methods such as Moving Window and Accelerated Moving Window can be applied in real time. Due to this reason, we assess the relation between change point detection delay and copula parameter in these two methods using simulated data. Subsequently, a series of data that imitate real financial market data is simulated, based on which, we make a comprehensive comparison of these four methods. Finally, the approach that performed best in detecting the change in copula is applied to historical data from Standard \& Poor 500 and Nasdaq index. The backtesting results of VaR and ES based on the dynamic copula illustrate that the dynamic model performs better than the static one.

The paper is organized as follows. The basic definitions, the goodness-of-fit test for detecting the change of copula family and parameter and the algorithms of four methods mentioned are introduced in Section \ref{sec:methodology}. In Section \ref{sec:Results}, we compare the performance of four methods in copula change based on the simulated data, we provide the empirical analysis by applying the best performing method and comparing dynamic and static modeling through backtesting. Conclusions are given in Section \ref{sec:conclusion}. Although, the focus in this paper is on the bivariate case in order to simplify the presentation, extension to higher dimensions is straightforward.

\section{Methodology}\label{sec:methodology}
In this section, we introduce some basic definitions, tests and the algorithms used to detect the change of dependence structure in financial data.

We consider a portfolio of $p$ (in this paper $p=2$) stocks. The one-period (e.g. one-day) loss can be written as a function of the one-period log-returns as: 
\begin{equation}\label{eqn:def:loss}
L_{t+1}=-(V_{t+1}-V_t)=-\sum_{i=1}^p\lambda_iS_t^i(\exp(X_{t+1})-1)
\end{equation}
where $V_t$ is the portfolio value that is known at time $t$ and $X_{t+1}=\ln S_{t+1}-\ln S_t=\ln \frac{ S_{t+1}}{S_t}$, where $S_{t}$ is the stock price at time $t$.



\subsection{Goodness-of-fit test}\label{subsec:gof}
It is assumed that the copula-based likelihood can be differentiated three times and the corresponding expectation can be computed, see Huang and Prokhorov \cite{huang2014goodness}. The rank-based goodness-of-fit test used here is related to the White \cite{white1982maximum} specification test and holds due to the information matrix equivalence theorem, which states that the Fisher information matrix is equal to the expected outer product of the score function or equal to minus the expected Hessian if the copula is correctly specified. 

Sklar's theorem \cite{sklar1959fonctions} describes how a multivariate distribution can be decomposed into a copula function and marginal functions. 

\begin{Theorem}\label{the:Sklar}
If $H$ is a joint distribution function of ($X_1,...,X_N$) with margins $F_1,...,F_N$, then there exists a copula $C$: $[0,1]^N\rightarrow [0,1]$ such that, for all $x_1,...,x_N$ in $\mathbb{R}=[-\infty, \infty]$,
\begin{equation}
H(x_1,...,x_N)=C(F_1(x_1),...,F_N(x_N)).
\end{equation}
Copula $C$ is unique if the margins are continuous.
\end{Theorem}
Assume the copula density exists, the joint density is
\begin{multline*}
h((x_1,...,x_N)=\left. \frac {\partial ^N C(u_1,...,u_N)}{\partial u_1,...,\partial u_N} \right|_{u_n=F_{n(x_n), n=1,...,N}} \prod _{n=1}^N f_n(x_n)=c(F_1 (x_1),...,F_N(x_N)) \prod _{n=1}^N f_n(x_n),
\end{multline*}
where $c$ is the copula density. Note that Sklar's theorem has a natural conditional extension by conditioning on information up to time $t$ (see Guegan and zhang \cite{guegan2010change}), which we omit here. 

The core of White's information matrix equivalence theorem \cite{white1982maximum} is that, under correct copula specification, 
$$
-\mathbb{H}(\theta _0 )=\mathbb{C}(\theta _0),
$$
where $\theta _0$ is the dependence parameter of copula under correct specification, $\mathbb{H}(\theta _0 )$ indicates the expected Hessian matrix of $\ln c_{\theta _0}$ and $\mathbb{C}(\theta _0)$ indicates the expected outer product of the corresponding score function and the expressions of $\mathbb{H}(\theta _0 )$  and $\mathbb{C}(\theta _0)$ are as below.
$$
\mathbb{H}(\theta _0 )=\mathbb{E}\nabla _{\theta _0}^2 \ln c_{\theta _0}(F_1(x_1),...,F_N(x_N))
$$
$$
\mathbb{C}(\theta _0)=\mathbb{E}\nabla _{\theta _0}\ln c_{\theta _0}(F_1(x_1),...,F_N(x_N))\nabla _{\theta _0}' \ln c_{\theta _0}(F_1(x_1),...,F_N(x_N)),
$$
Where ``$\nabla _{\theta _0}$'' represents derivatives regarding $\theta _0$ and expectations are related to the true distribution $H$.

Now the hypothesis of the copula misspecification test is, see Huang and Prokhorov \cite{huang2014goodness}:
\begin{equation}\label{eqn:def:hyp}
H_0 : \mathbb{H}(\theta _0)+\mathbb{C}(\theta _0)=0 \qquad versus \qquad
H_1 :\mathbb{H}(\theta _0)+\mathbb{C}(\theta _0)\neq 0
\end{equation}
In practice, $\theta _0$ is replaced by a consistent estimator $\widehat{\theta}$ and the marginal distribution function $F_N$ is replaced by the empirical distribution function $\widehat{F_N}$. 

The empirical distribution function is 
$$
\widehat{F_n}(s)=T^{-1}\sum _{t=1}^T I\{x_{nt} \leq s\},
$$
where $T$ is the length of each margin, $I\{ \cdot \} $ is the indicator function and $s$ represents the observed value of $x_n$. Then, $\widehat{\theta}$ can be calculated as 
$$
\max _{\theta _0} \sum _{t=1}^T \ln  c_{\theta _0}(\widehat{F_1}(x_{1t}),...,\widehat{F_N}(x_{Nt})).
$$

As for $\mathbb{H}$ and $\mathbb{C}$, they are replaced by the sample averages $\overline{\mathbb{H}}$ and $\overline{\mathbb{C}}$ respectively, which are computed from empirical estimation $\widehat{\mathbb{H}}$ and $\widehat{\mathbb{C}}$.
$$
\widehat{\mathbb{H}_t} (\theta_0)=\nabla_{\theta_0}^2 \ln c_{\theta_0}(\widehat{F_1}(x_{1t}),...,\widehat{F_N}(x_{Nt}))
$$
$$
\widehat{\mathbb{C}_t} (\theta_0)=\nabla_{\theta_0} \ln c_{\theta_0}(\widehat{F_1}(x_{1t}),...,\widehat{F_N}(x_{Nt}))\nabla_{\theta_0}'  \ln c_{\theta_0}(\widehat{F_1}(x_{1t}),...,\widehat{F_N}(x_{Nt})).
$$
Then, $\overline{\mathbb{H}}$ and $\overline{\mathbb{C}}$ can be expressed as 
$$
\overline{\mathbb{H}}(\theta_0)=T^{-1}\sum_{t=1}^T\widehat{\mathbb{H}_t}(\theta_0)
$$
$$
\overline{\mathbb{C}}(\theta_0)=T^{-1}\sum_{t=1}^T\widehat{\mathbb{C}_t}(\theta_0).
$$

White \cite{white1982maximum} defined
$$
\overline{D}_{\theta_0}\equiv \overline{D}(\theta_0)\equiv T^{-1}\sum_{t=1}^T\widehat{d_t}(\theta_0)
$$
$$
\widehat{d_t}(\theta_0)=vech(\widehat{\mathbb{H}_t} (\theta_0)+\widehat{\mathbb{C}_t} (\theta_0)),
$$
Where $vech$ represents the vertical vectorization of the lower triangle of a matrix. Here, $\widehat{d_t}$ based on the empirical distribution.

Now the test statistic can be constructed as follows, see Huang and Prokhorov \cite{huang2014goodness}.
\begin{Proposition}\label{GOF}
Under correct copula specification and suitable regularity conditions, the information matrix test statistic is 
\begin{equation}\label{eqn:def:GOF}
\mathcal{F}=T \overline{D_{\widehat{\theta}}'} V_{\theta_0}^{-1}\overline{D_{\widehat{\theta}}}
\end{equation}
where $V_{\theta_0}$ is given below.
\begin{dmath*}
V_{\theta_0}=E\{d_t(\theta_0)+\nabla D_{\theta_0} B^{-1} [\nabla _{\theta} \ln c (F_{1t},F_{2t},...,F_{Nt}; \theta _0)+\sum _{n=1}^N W_n(F_{nt})]+\sum _{n=1}^N M_n(F_{nt})\} \times \{d_t(\theta_0)+\nabla D_{\theta_0} B^{-1} [\nabla _{\theta} \ln c (F_{1t},F_{2t},...,F_{Nt}; \theta _0)+\sum _{n=1}^N W_n(F_{nt})]+\sum _{n=1}^N M_n(F_{nt})\}',
\end{dmath*}
where $d_t$ dependents on unknown margins and 
\begin{equation*}
D_{\theta _0}=\mathbb{E}d_t(\theta _0).
\end{equation*}
For $n=1,2,...,N$,
\begin{dmath*}
W_n(F_{nt})=\int _0^1 \int _0^1\cdot \cdot \cdot \int _0^1 [I\{F_{nt} \leq u_n\}-u_n] \nabla _{\theta ,u_n}^2 \ln c(u_1,u_2,...,u_N;\theta _0)c(u_1,u_2,...,u_N;\theta _0)du_1du_2 \cdot \cdot \cdot du_N,
\end{dmath*}
and
\begin{dmath*}
M_n(F_{nt})=\int _0^1 \int _0^1\cdot \cdot \cdot \int _0^1 [I\{F_{nt} \leq u_n\}-u_n] \nabla _{u_n} vech[\nabla _{\theta}^2\ln c(u_1,u_2,...,u_N;\theta _0)+\nabla _{\theta} \ln c(u_1,u_2,...,u_N;\theta _0)\nabla _{\theta}'\ln c(u_1,u_2,...,u_N;\theta _0)]c(u_1,u_2,...,u_N;\theta _0)du_1du_2 \cdot \cdot \cdot du_N.
\end{dmath*}
\end{Proposition}

The statistic $\mathcal{F}$ is distributed asymptotically as $\chi_{p(p+1)/2}^2$, where $p$ is the dimension of $\theta _0$. Hence, for Student-t copula, $\mathcal{F}$ follows a Chi-square distribution with three degrees ($p=2$) while for Gaussian and other copulas with one parameter, $\mathcal{F}$ tends to a Chi-square distribution with one degrees. Based on this test, we can check whether the copula is not changing during a period.

\subsection{Change point detection}\label{subsec:detect}
Rejection of the null hypothesis $H_0$ means there is enough evidence that the copula does not remain constant in the section of data analyzed. That means either the family or the parameters have changed and we should model the copula dynamically. We introduce 2 new methods for change point detection, namely Bottom-up and Accelerated Moving Window and compare their performance to the existing Binary Segmentation and Moving Window methods. Binary segmentation and Bottom-up can only be used retrospectively, i.e. to fit copulas dynamically and identify change points in past data. The other two are useful for "real-time" change point detection. We describe each method in detail in the remaining part of this section starting from the existing ones.
\subsubsection{Binary Segmentation}\label{subsec:bs}
Binary Segmentation method is a commonly used method when detecting the change of the dependence structure, and only one point can be detected each time. We can use the following algorithm.

\begin{Algorithm}\label{def:BS} 
\mbox{}\par
\begin{enumerate}
\item[(1)] Fit the best copula for the whole sample according to Akaike Information criterion (AIC).
\item[(2)] Apply the goodness-of-fit test and obtain the test statistic.
\item[(3)] While the null hypothesis ($H_0$: Copula does not change) is rejected:
\begin{enumerate}
\item Split the sample into two subsamples.
\item Fit the best copula for each subsample (AIC).
\item Apply the goodness-of-fit test for each subsample and obtain the test statistic.
\end{enumerate}
\end{enumerate}
\end{Algorithm}

\subsubsection{Moving Window}\label{subsec:mw}
The Moving Window method proposed by Caillault and Guegan \cite{guegan2009forecasting} starts from a fixed interval and then moves forward until the whole data is covered. In each interval, Caillault and Guegan \cite{guegan2009forecasting} fitted the  best copula while in this paper, we test the goodness-of-fit instead of just choosing the best copula. We choose the rolling-window size as $N$ points and move this window by $K$ points every time in order to maintain a good power of test. Apart from the number of data, the power of test also depends on parameters like the correlation so one must carefully select $K$ and $N$. In this paper we use $N=500$ and $K=120$. The algorithm is as follows:
\begin{Algorithm}\label{def:MW} 
\mbox{}\par
\begin{enumerate}
\item[(1)] Take $N$ initial data.
\item[(2)] Fit the copula model according to AIC.
\item[(3)] Apply the goodness-of-fit test and obtain the test statistic.
\item[(4)] While $H_0$ is rejected (Test statistic $< \chi_{p(p+1)/2,\alpha}^2$ for the White test):
\begin{enumerate}
\item Move the $N$-rolling window forward by $K$ points. 
\item Apply the goodness-of-fit test and obtain the test statistic.
\end{enumerate}
\item[(5)] When (Test statistic $> \chi_{p(p+1)/2,0.95}^2$)
\begin{enumerate}
\item Drop all old observations except the last $K$ points from the change point.
\item Set a new $N$-rolling window.
\item Choose the best copula (AIC).
\item Apply the goodness-of-fit test.
\item Back to step (4).
\end{enumerate}
\end{enumerate}
\end{Algorithm}

\subsubsection{Accelerated Moving Window}\label{subsec:amw}
The main idea of the Accelerated Moving Window method is monitoring the movement of the test statistic that produced by the goodness-of-fit test. Here we take advantage of the observation that the test statistic monotonically increases when data that come from a different model start to be added to the window. Once the statistic exceeds the warning limit line (WLL), there is a  signal that the copula has changed. Then we set a new window from the point of WLL and monitor the movement of the statistic until it crosses the control limit line (CLL). The warning limit and control limit are determined by two confidence levels $\alpha_w$ and $\alpha_c$ respectively where $\alpha_w < \alpha_c$. The hope is that by dropping old data after the WLL point, the Accelerated Moving Window will detect the change point earlier than the Moving Window. Here, we take $\alpha_w=0.85$ for the warning limit ($\chi_{3,0.85}^2=5.32$ and  $\chi_{1,0.85}^2=2.07$  for the Student-t copula and the copula with one parameter respectively). For the CLL, we consider $\alpha_c=0.95$ (Student-t copula takes  $\chi_{3,0.95}^2=7.81$ and other copulas with one parameter take $\chi_{1,0.95}^2=3.84$). The algorithm is as follows.
\begin{Algorithm}\label{def:AIMW} 
\mbox{}\par
\begin{enumerate}
\item[(1)] Take $N_{min}$ initial data (in the examples we use $N_{min}=200$).
\item[(2)] Fit the copula model according to AIC criteria.
\item[(3)] Apply the goodness-of-fit test and obtain the test statistic.
\item[(4)] While (Test statistic $< \chi_{p(p+1)/2,\alpha_w}^2$ )
\begin{enumerate}
\item While the length of the window $N$ is less than $L$ (here $L=500$)
\begin{enumerate}
\item Add $D$ new points (here $D=50$). 
\item Apply the goodness-of-fit test and obtain the test statistic.
\end{enumerate}
\item Else ($N=L$)
\begin{enumerate}
\item Move the $N$-rolling window forward $K$ points.
\item Apply the goodness-of-fit test and obtain the test statistic.
\end{enumerate}
\end{enumerate}
\item[(5)] When (Test statistic $> \chi_{p(p+1)/2,\alpha_w}^2$)
\begin{enumerate}
\item Drop all old observations before current point except the last $K$ points.
\item Set a new window with $N_{min}$ points.
\item Fit the copula model according to AIC criteria.
\item Apply the goodness-of-fit test and obtain the test statistic.
\end{enumerate}
\item[(6)] While (Test statistic $< \chi_{p(p+1)/2,\alpha_c}^2$)
\begin{enumerate}
\item While ($N<L$) 
\begin{enumerate}
\item Add $D$ points. 
\item Apply the goodness-of-fit test and obtain the test statistic.
\end{enumerate}
\item Else ($N=L$)
\begin{enumerate}
\item Move the $N$-rolling window forward $K$ points.
\item Apply the goodness-of-fit test and obtain the test statistic.
\end{enumerate}
\end{enumerate}
\item[(7)] When (Test statistic $> \chi_{p(p+1)/2,\alpha_c}^2$)
\begin{enumerate}
\item Drop all old observations before the change point except the last $K$ points .
\item Set a new window with $N_{min}$ points.
\item Choose the best copula (AIC).
\item Apply the goodness-of-fit test and obtain the test statistic. 
\item Back to step (4).
\end{enumerate}
\end{enumerate}
\end{Algorithm}

\subsubsection{Bottom-up method}\label{subsec:bu}
Inspired by the ``Tail-greedy bottom-up decompositions'' of Fryzlewicz \cite{fryzlewicz2016tail}, the Bottom-up method first splits the whole data into multiple sub-segments with equal length $N^*_{min}$(in principle, this $N^*_{min}$ should be the same as above to justify fairness, while the truth is that the bottom-up feature of this method decides its own suitable size) and then selects the best copula for each sub segment. It is important to note that the interval should be small enough not to contain any change point (or at least cause a delay in detection) and large enough to ensure enough data are present to fit the copula. After that, the merging process proceeds layer by layer if three conditions are satisfied. First, contiguous sub-segments belong to the same copula family. Second, the copula family that obtained by combining two sub-segments remains unchanged. Third, the statistic of goodness-of-fit test on the pooled data should be below the CLL, which indicates that there is no change point in the merged segment. Once all the possible sub-segments have been merged, the change points of copula can be found. As for the minimum size of each sub-segment, here we chose $N^*_{min}=100$ points, or fewer if the goodness-of-fit test indicates the variant in the copula. We can use the following algorithm. 
\begin{Algorithm}\label{def:BU} 
\mbox{}\par
\begin{enumerate}
\item[(1)] Divide the whole data into multiple sub-segments with $N^*_{min}$ points in each segment.
\item[(2)] Fit the best the copula for each segment according to AIC criterion.
\item[(3)] Apply the goodness-of-fit test and obtain the test statistic.
\item[(4)] While (Any test statistic $> \chi_{p(p+1)/2,\alpha_c}^2$)
\begin{enumerate}
\item Adjust the length of the sub-segments.
\item Back to (2).
\end{enumerate}
\item[(5)] While (The family of the contiguous sub-segments is the same)
\begin{enumerate}
\item Merge the contiguous sub segments.
\item Fit the best copula for the merged segment and obtain a new copula family (AIC).
\item Apply the goodness-of-fit test to the merged segment and obtain the test statistic.
\item If (The new copula family is consistent with the family of the contiguous sub segments \& Test statistic $< \chi_{p(p+1)/2,\alpha_c}^2$)
\begin{enumerate}
\item Keep merged.
\end{enumerate}
\item Else
\begin{enumerate}
\item Keep the sub segments.
\end{enumerate}
\end{enumerate}
\end{enumerate}
\end{Algorithm}

\begin{figure}[H]
\centering
\includegraphics[width=16cm]{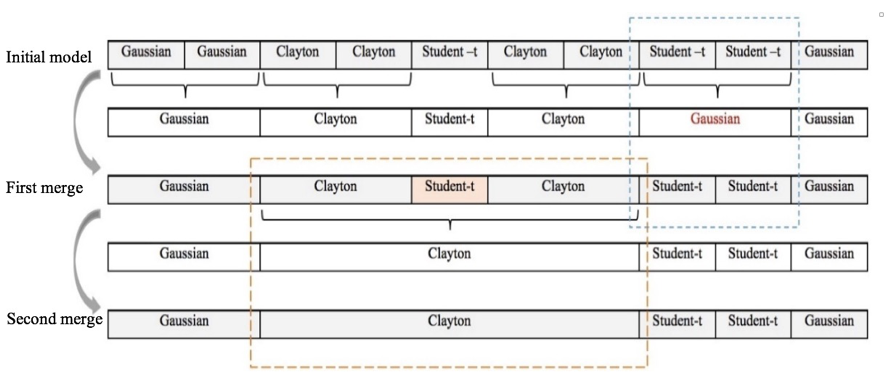}
\caption{A demonstration of Bottom-up method.}
\label{Fig.bu}
\end{figure}

Figure \ref{Fig.bu} is a demonstration of the Bottom-up method. The copulas from ``initial model'' are merged successfully such as Gaussian copula and Clayton copula displayed in ``First merge'', while the Student-t copula fails the merger since the merged copula is Gaussian copula, shown in the blue grid. The orange grid represents a special case when dealing with the merger. If the Student-t copula has a slight impact on the combination of the Clayton and Student-t copula, we can merge the data and represent as one Clayton copula to avoid overfitting.

\section{Results-Comparison of methods}\label{sec:Results}
In this paper, the performance of the methods introduced in Section \ref{subsec:detect} are compared based on the simulated data. We compare Moving Window and Accelerated Moving Window since they can be applied in real time. Binary Segmentation and Bottom-up method will form another comparison pair as they can only be applied retrospectively. After that, the best performed method is applied to model the dependence structure of Standard \& Poor 500 (S\&P 500) and Nasdaq indices.
\subsection{Simulated Data}\label{subsec:simulated}
\subsubsection{Family change in simulated data}\label{subsec:family}
\begin{figure}[H]
\centering
\includegraphics[width=10cm]{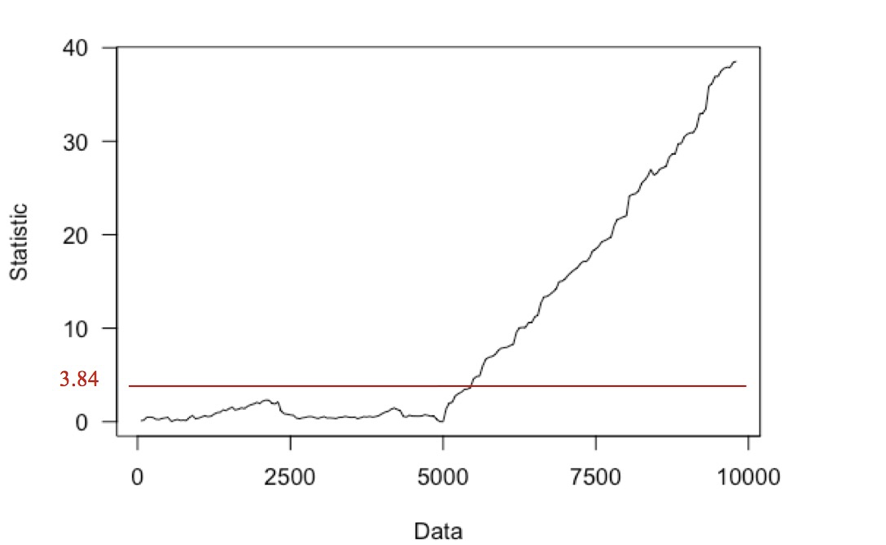}
\caption{The movement of statistics that produced by the goodness-of-fit test when the combination is Gaussian copula (0.5) and Student-t copula (0.5, 2.2).}
\label{Fig.3}
\end{figure}

To replicate the results, we generate 10000 random data from two copula families with equal weight and same parameters, then apply four methods respectively. Figure \ref{Fig.3} shows the trend of the statistic that was produced by the goodness-of-fit test of the combination between Gaussian copula (0.5) and Student-t copula (0.5, 2.2) when applying Accelerated Moving Window method. There is a significant jump after 5000 points and the test statistic obviously exceeds the critical value at confidence level $\alpha=95 \%$, that is 3.84 ($\chi _{1,0.95}^2$). It is evident that the copula family has changed.

Table 1 illustrates the results of different combinations of copula family when applying Moving Window and Accelerated Moving Window methods. We only focus on Gaussian copula, Student-t copula and Clayton copula. The number in the bracket represents the parameter of the copula, which denotes the strength of dependency and ``True change point'' representing the real change point in the simulated data. ``Detected change point'' denotes the change point that detected by different methods and ``Delay'' means the corresponding delay of ``Detected change point'' compared with ``True change point''. ``New copula'' represents the copula that after the change point and it differs from the copula before the change point. Conclusions of comparing different methods are summarized below:
\begin{enumerate}
\item Both Moving Window and Accelerated Moving Window are able to detect the change points in all the combinations of copula family.
\item Accelerated Moving Window responds to the change of copula much faster than Moving Window method when the combinations are Clayton/Student-t, Gaussian/Clayton, Clayton/Gaussian and Student-t/Clayton. Especially for the combination of Clayton and Gaussian copula, Accelerated Moving Window only spends 1.2 years (299 points) to react to the change while Moving Window takes almost 19 years (4739 points). While 1.2 years remains impractical for real life use, it is a vast improvement and in specific applications with fine tuning  of parameters has the potential to become a real time warning system. 
\item Accelerated Moving Window performs less well in the combinations of Gaussian/Student-t as well as Student-t/Gaussian, and slower than Moving Window roughly by one month (19 points) and 3.6 months (79 points) respectively. 
\end{enumerate}
\begin{table}[htbp]
\small
\centering
    \caption{\label{tab:1}Performance of Moving Window method and Accelerated Moving Window method}
      \begin{threeparttable}
        \begin{tabular}{c|ccccc}
            \hline
            Combination of copula & Approach &  \tabincell{c}{True \\ change \\ point} &  \tabincell{c}{Detected \\ change \\ point} & Delay & New copula  \\
             \hline
             \multirow{3}*{\tabincell{c}{ Gaussian (0.5) \\  Student-t  (0.5, 2.2)}} 
             &  Moving Window & 5001 & 5300 & 299 & Student-t (0.52, 2) \\ 
          
             ~ & \tabincell{c}{Accelerated \\ Moving Window} & 5001 & 5319 & 318 & Student-t (0.47, 2)  \\ 
             \hline
            \multirow{3}*{\tabincell{c}{Clayton (0.5) \\ Student-t  (0.5, 2.2)}}  
             &  Moving Window & 5001 & 5300 & 299 & Student-t (0.52, 2) \\ 
             
             ~ & \tabincell{c}{Accelerated  \\ Moving Window} & 5001 & 5180 & 179 & Student-t (0.41, 2)  \\ 
              \hline
            \multirow{3}*{\tabincell{c}{Gaussian (0.5) \\   Clayton (0.5)}} 
             &  Moving Window & 5001 & 6620 & 1619 & Clayton (0.43) \\ 
            
             ~ & \tabincell{c}{Accelerated \\ Moving Window} & 5001 & 5799 & 798 & Clayton (0.44)  \\
              \hline
            \multirow{3}*{\tabincell{c}{Clayton (0.5) \\  Gaussian (0.5)}} 
             &  Moving Window & 5001 & 9740 & 4739 & Gaussian (0.57)  \\ 
             
             ~ & \tabincell{c}{Accelerated \\ Moving Window} & 5001 & 5300 & 299 & Gaussian (0.47) \\
              \hline
            \multirow{3}*{\tabincell{c}{Student-t  (0.5,2.2) \\ Gaussian (0.5)}}
             &  Moving Window & 5001 & 5300 & 299 & Gaussian (0.47)  \\ 
             ~ & \tabincell{c}{Accelerated \\ Moving Window} & 5001 & 5379 & 378 & Gaussian (0.45) \\
              \hline
           \multirow{3}*{\tabincell{c}{Student-t  (0.5, 2.2) \\ Clayton (0.5)}} 
             &  Moving Window & 5001 & 5300 & 299 & Clayton (0.45)  \\ 
             
             ~ & \tabincell{c}{Accelerated \\ Moving Window} & 5001 & 5259 & 258 & Clayton (0.44) \\
              \hline    
        \end{tabular}
         \begin{tablenotes}
        \footnotesize
        \item Figures in brackets are copula parameters. For Student-t copula, the first parameter is correlation and the second one is degree of freedom
. ``NA'' represents that no change point is detected.
      \end{tablenotes}
    \end{threeparttable}
\end{table}
\begin{table}[htbp]
\small
\centering
    \caption{\label{tab:2}Performance of Binary Segmentation and Bottom-up method}
      \begin{threeparttable}
        \begin{tabular}{c|ccccc}
            \hline
            Combination of copula & Approach &  \tabincell{c}{True \\ change \\ point} &  \tabincell{c}{Detected \\ change \\ point} & \tabincell{c} {Distance from \\true\\ change point} & New copula  \\
             \hline
             \multirow{2}*{\tabincell{c}{ Gaussian (0.5) \\ Student-t (0.5, 2.2)}} 
             &  Binary Segmentation & 4501 & NA & NA & NA \\ 
             ~ &  Bottom-up & 4501 & 4401 & -100 & Student-t (0.46, 6.53) \\ 
             \hline
            \multirow{2}*{\tabincell{c}{Clayton (0.5) \\ Student-t (0.5, 2.2)}} 
             &  Binary Segmentation & 4501 & 5001 & 500 & Student-t (0.49, 4.67) \\ 
             ~ &  Bottom-up & 4501 & 4601 & 100 & Student-t (0.53, 5.01) \\
              \hline
            \multirow{2}*{\tabincell{c}{Gaussian (0.5) \\  Clayton (0.5)}} 
             &  Binary Segmentation &4501 & NA & NA & NA \\ 
             ~ &  Bottom-up & 4501 & 4401 & -100 & Clayton (0.62) \\
              \hline
            \multirow{2}*{\tabincell{c}{ Clayton (0.5) \\  Gaussian (0.5)}} 
             &  Binary Segmentation & 4501 & NA & NA & NA \\ 
             ~ &  Bottom-up & 4501 & 4601 & 100 & Clayton (0.54) \\
              \hline
            \multirow{2}*{\tabincell{c}{Student-t (0.5, 2.2) \\ Gaussian (0.5)}} 
             &  Binary Segmentation & 4501 & NA & NA & NA \\ 
             ~ &  Bottom-up & 4501 & 4601 & 100 & Gaussian (0.54) \\
              \hline
            \multirow{2}*{\tabincell{c}{Student-t (0.5, 2.2) \\ Clayton (0.5)}} 
              &  Binary Segmentation & 4501 & 5001 & 500 & Clayton (0.48) \\ 
             ~ &  Bottom-up & 4501 & 4601 & 100 & Clayton (0.61) \\
              \hline    
        \end{tabular}
         \begin{tablenotes}
        \footnotesize
        \item Figures in brackets are copula parameters. For Student-t copula, the first parameter is correlation and the second one is degree of freedom
. ``NA'' represents that no change point is detected.
      \end{tablenotes}
    \end{threeparttable}
\end{table}

Table 2 displays the performance of Binary Segmentation and Bottom-up method in detecting the change of copula family. There is no doubt that Bottom-up method is outstanding compared with Binary Segmentation since it is sensitive to the family change and even detects the change points early than the true change points. However, Binary Segmentation displays disappointing performance and fails to identify more than 50\% of changes in copula family change.

\subsubsection{The relation of delay and parameter change in simulated data}\label{subsec:parameter}
\begin{figure}[H]
\centering
\subfigure[]{
\label{Fig.sub.11}
\includegraphics[width=7.5cm]{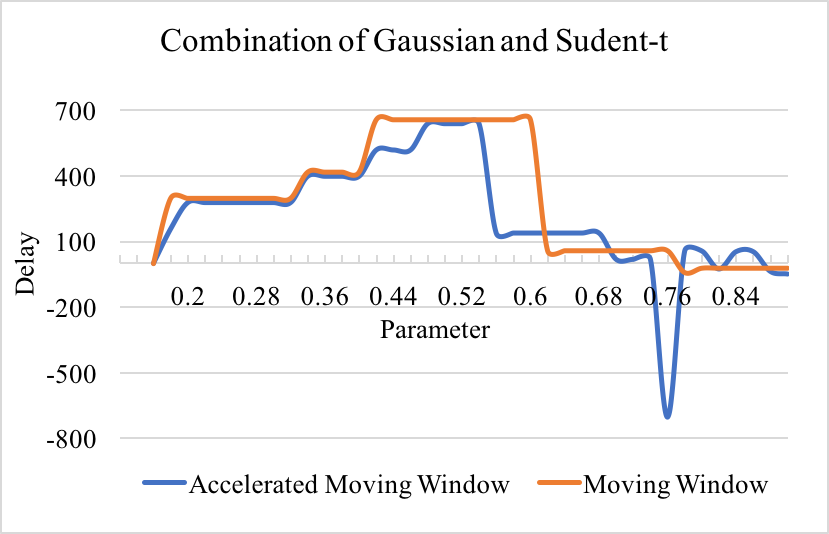}}
\subfigure[]{
\label{Fig.sub.21}
\includegraphics[width=7.5cm]{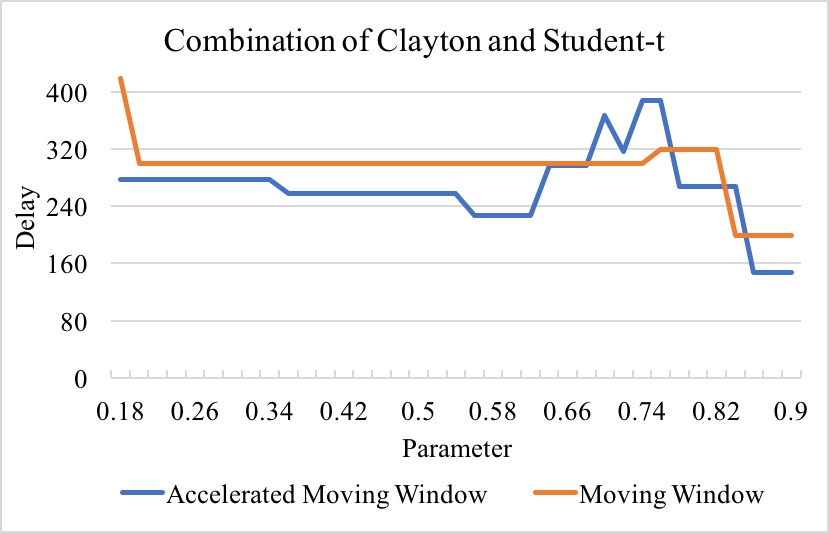}}
\subfigure[]{
\label{Fig.sub.31}
\includegraphics[width=7.5cm]{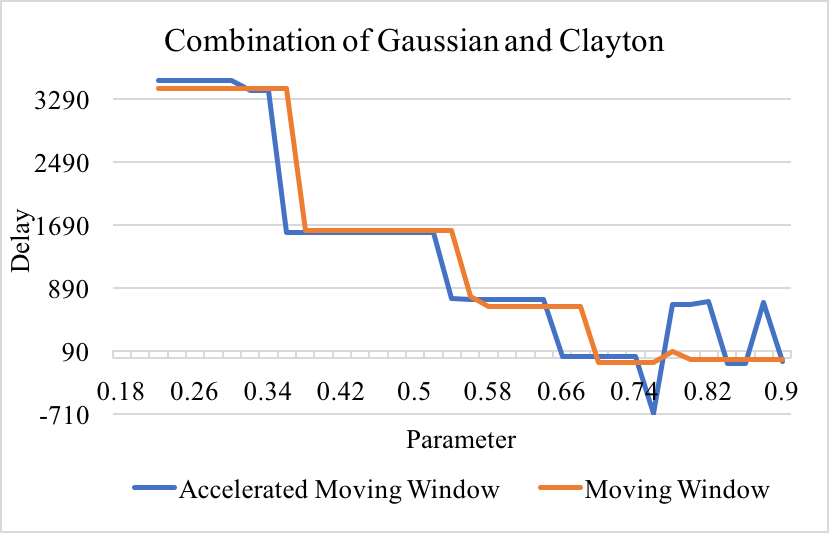}}
\subfigure[]{
\label{Fig.sub.41}
\includegraphics[width=7.5cm]{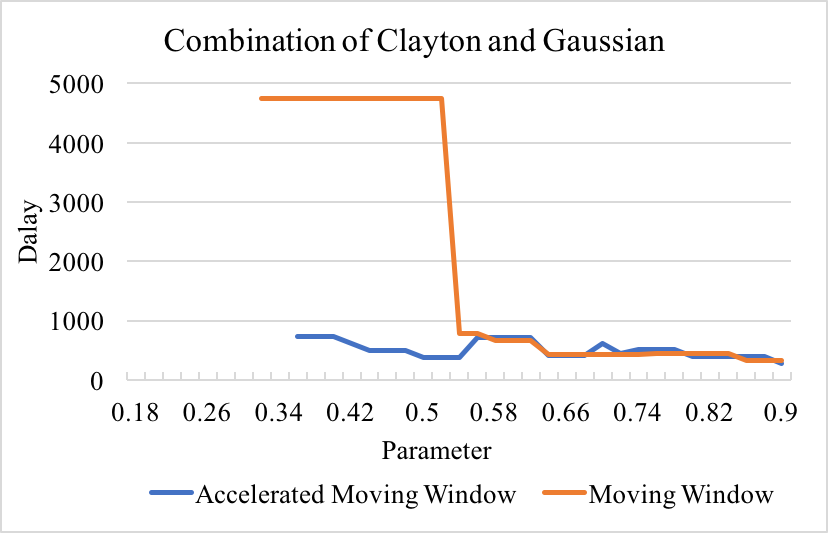}}

\caption{The relation of delay and parameter change.}
\label{Fig.delay}
\end{figure}

In this section, we study the relation of delay and parameter change by two methods, which are Moving Window and Accelerated Moving Window methods. We still use seed (626) and generate 10000 random data from two copula families but with different parameters. Figure \ref{Fig.delay} shows the results and we can draw several conclusions from it. 
\begin{enumerate}
\item For the combination of Gaussian and Student-t copula, both Accelerated Moving Window and Moving Window are more sensitive to the change when the correlation is either lower than 0.4 or higher than 0.6. For the other three combinations, the increment of copula parameter contributes to lower delay except for some fluctuation, which means these two methods tend to respond quickly to higher dependence in the copula change.
\item For the combinations of Gaussian and Student-t copula as well as Student-t and Gaussian copula, Accelerated Moving Window method performs better than Moving Window method when the parameters are lower than 0.6, while shows some fluctuations when the parameters are higher than 0.6. In general, Accelerated Moving Window outperforms than Moving Window. 
\item For the combination of Gaussian and Clayton copula, the results of Accelerated Moving Window and Moving Window are similar to each other.
\item For the combination of Clayton and Gaussian copula, it is evident that Accelerated Moving Window method is prevailing when the parameters are less than 0.5 since it is faster than Moving Window method about 4000 points, which are equivalent to about 16 years.  
\end{enumerate}

\subsubsection{More complicated situations}\label{subsec:complicated}
\begin{figure}[H]
\centering
\includegraphics[width=16cm]{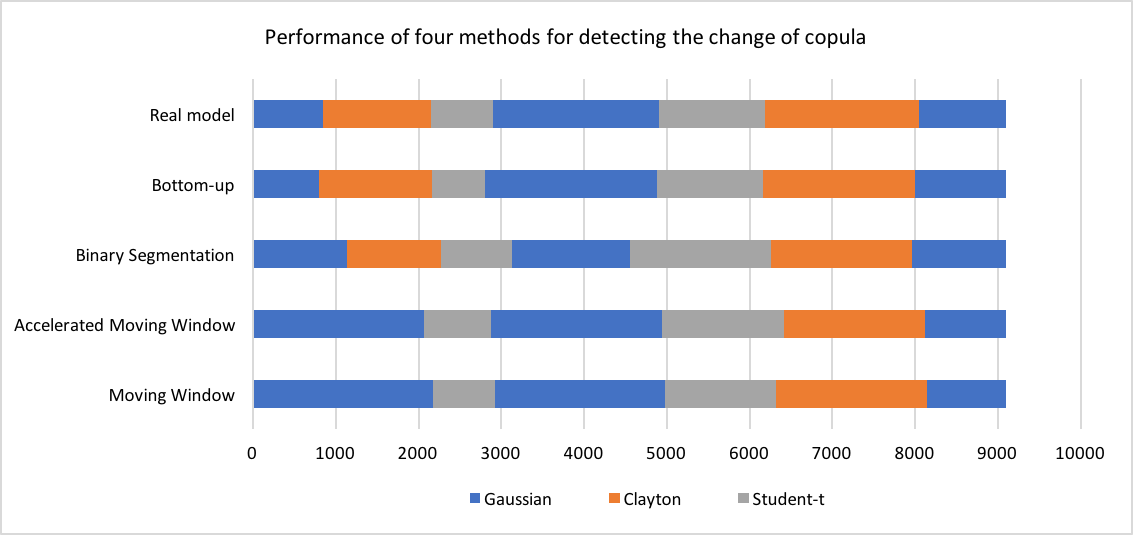}
\caption{The performance of the four methods for detecting the change of copula in a complicated simulation.}
\label{Fig.5}
\end{figure}

Based on the results above, it is worth to observe the performance of four methods in detecting the copula change in a more complicated case. Here, we attempt to simulate a series data that close to the behavior of the real financial market. Figure \ref{Fig.5} shows the real change in our simulated data and the copula change that detected by four methods respectively. ``Real model'' contains 9100 pairs of data and consists of Gaussian, Student-t and Clayton copula. The conclusions are as follows.
\begin{enumerate}
\item Bottom-up method is the best performing method since it is able to detect all the copula changes with small deviation from the real change point. However, this method bears the weakness that it is not possible to be applied in the real time.
\item Binary Segmentation performs well and detects all the copula families although appears to be inaccurate. It cannot be implemented in real time as well.
\item Both Accelerated Moving Window and Moving Window fail to detect the Clayton copula change between points 850 and 2150. Accelerated Moving Window responds to the change faster than Moving Window for all the copula families except the Clayton copula (between point 6180 and 8040), but is still less accurate than Bottom-up method.
\end{enumerate}

After a series of comparisons, the Bottom-up method seems to be the most accurate method. Accelerated Moving Window definitely outperforms than Moving Window and is able to accelerate the detection of the change point in most cases,  while still performs worse than Bottom-up method. Therefore, the Bottom-up method is preferred method to be applied to Standard \& Poor 500 (S\&P 500) and Nasdaq indices to evaluate the effectiveness.
\subsection{Financial Time Series}\label{subsec:financial}
To study how the dependency of Standard \& Poor 500 (S\&P 500) and Nasdaq indices change during an long period, we choose the data from 4 January, 2005 to 31 December, 2015, which contains 2768 daily observations. 

\begin{figure}[H]
\centering
\includegraphics[width=12.5cm]{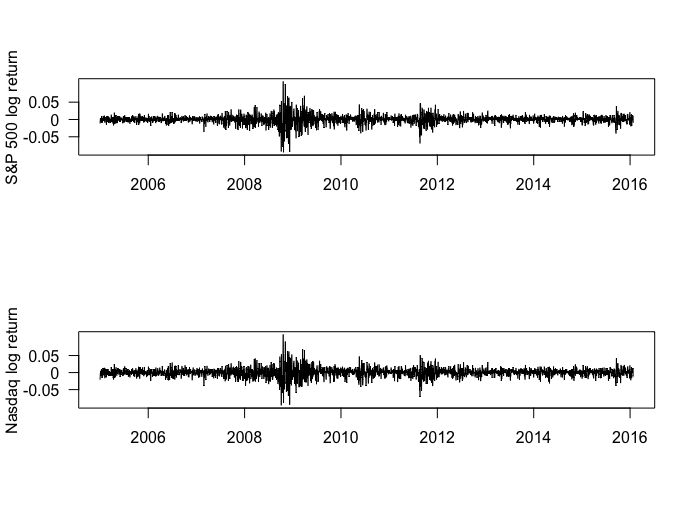}
\caption{Log-returns of S\&P 500 and Nasdaq indices.}
\label{Fig.6}
\end{figure}

The log-returns of two indices are displayed in Figure \ref{Fig.6}, from which we can see that the time series of these log-returns resemble each other and almost move in the same direction. It is evident that the peak of two assets occurs simultaneously, especially during the 2008 financial crisis. 

\subsubsection{Marginal model}\label{subsec:marginal}
Considering the ``stylized facts'' \cite{mcneil2015quantitative} of the real financial market such as changing volatility and asymmetry. We fit the GARCH (2,1) model with Normal innovation to each log-return. Here, $r_{i,t}(i=1,2)$ represents the daily log-returns of S\&P 500 and Nasdaq indices respectively. The model is defined as:
\begin{gather*}
r_{i,t}=\mu _i+\xi _{i,t},\\
\xi _{i,t}=\sigma _{i,t}\varepsilon _{i,t},\\
\sigma _{i,t}^2=\alpha _{i,0}+\alpha _{i,1}\varepsilon _{i,t-1}^2+\alpha _{i,2}\varepsilon _{i,t-2}^2+\beta _{i,1}\sigma _{i,t-1}^2,\\
\varepsilon _{i,t}|\varphi _{i,t} \sim N(0,1),
\end{gather*}
where $\mu _i$ is the drift, and $\alpha _{i,0},\alpha _{i,1},\alpha _{i,2},\beta _{i,1}$ are parameters in $\mathbb{R}$. Table \ref{tab:3} exhibits the estimated results of GARCH (2, 1) model by the maximum log-likelihood method.

\begin{table}[htbp]
\small
\centering
    \caption{\label{tab:3}Estimation of GARCH (2, 1) model}
      \begin{threeparttable}
        \begin{tabular}{p{3.2cm}<{\centering}p{5.2cm}<{\centering}p{5.2cm}<{\centering}}
            \hline
            Parameter &  S\&P 500 & NASDAQ  \\
             \hline
             $\mu$ & 5.833e-04 (1.506e-04) & 6.738e-04 (1.819e-04) \\
             $\alpha _0$ & 3.218e-06 (5.782e-07) & 3.759e-06 (7.746e-07) \\
             $\alpha _1$ & 2.726e-02 (1.651e-02) & 3.357e-02 (1.734e-02) \\
             $\alpha _2$ & 1.120e-01 (2.302e-02) & 7.399e-02 (2.185e-02) \\
             $\beta _1$ & 8.335e-01 (1.749e-02) & 8.656e-01 (1.584e-02) \\
             \hline    
        \end{tabular}
         \begin{tablenotes}
        \footnotesize
        \item Figures in bracket are standard error. 
      \end{tablenotes}
    \end{threeparttable}
\end{table}

\subsubsection{Detect the family and parameter change of copula}\label{subsec:change}
To model the dependence structure between two indices, first, we fit the copula for the standard residual pairs ($\varepsilon _{1,t},\varepsilon _{2,t}$) derived from GARCH (2, 1) model and then choose the best copula according to AIC criterion. The copula family set contains Student-t, Gaussian and Clayton copula. From Table \ref{tab:4}, we observe that Student-t is the best-fitted copula which has the smallest AIC.

\begin{table}[htbp]
\small
\centering
    \caption{\label{tab:4}Copula fitting results}
      \begin{threeparttable}
        \begin{tabular}{p{2.2cm}<{\centering}p{2.7cm}<{\centering}p{2.7cm}<{\centering}p{2.2cm}<{\centering}p{2.2cm}<{\centering}}
            \hline
            Copula & Parameter & Log-likelihood & AIC & Convergence  \\
             \hline
             \multirow{2}*{\tabincell{c}{ Student-t }} 
             &  0.94 (0.003)& \multirow{2}*{3202.49} & \multirow{2}*{-6400.17} & \multirow{2}*{T} \\ 
             ~ &  2.89 &  &  & \\             
             Gaussian & 0.9439 (0.002) & 3061 & -6119.694 & T \\
             Clayton & 5.045 (0.137) & 2670 & -5337.857 & T \\
             \hline    
        \end{tabular}
         \begin{tablenotes}
        \footnotesize
        \item Figures in bracket are standard error. For Student-t copula, the first parameter is correlation and the second one is degree of freedom. ``T'' denotes ``True'' and ``F'' denotes ``False''.
      \end{tablenotes}
    \end{threeparttable}
\end{table}
Now we have fitted the copula model for the data sample under the assumption that the dependency is static. Next, the rank-based goodness-of-fit test that proposed by Huang and Prokhorov \cite{huang2014goodness} is applied to test the constancy of Student-t copula specified in Table \ref{tab:4}. The test statistic is calculated based on the equation:
\begin{equation*}\label{eqn:def:GOF}
\mathcal{F}=T \overline{D_{\widehat{\theta}}'} V_{\theta_0}^{-1}\overline{D_{\widehat{\theta}}}
\end{equation*}
The result is $\mathcal{F}=$34.4843, which is far more than $\chi _{3,0.95}^2=$7.81. In addition, the corresponding p-value equals to 1.565551e-07. Hence, we reject the null hypothesis that the copula remains unchanged and model the dependency in a dynamic way.

\begin{figure}[H]
\centering
\includegraphics[width=16cm]{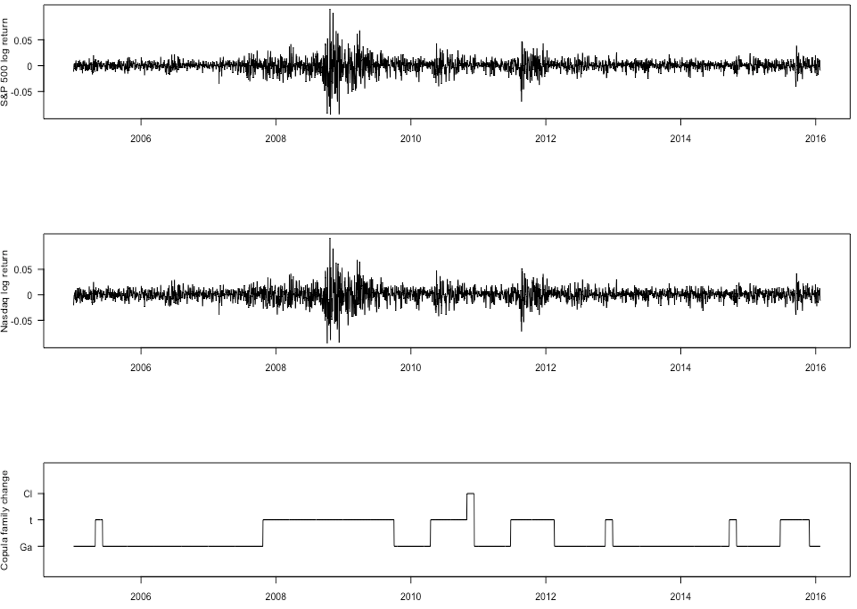}
\caption{Detection of copula family change. ``Ga'' represents the Gaussian copula, ``t'' represents the Student-t copula, ``Cl'' represents the Clayton copula.}
\label{Fig.7}
\end{figure}

In the next step, we apply Bottom-up method to detect the change of both copula family and parameter. Recall the algorithm \ref{def:BU} introduced in section \ref{subsec:bu} we first split the whole data into several segments with equal length. Then, choose the appropriate copula for each segment and apply the goodness-of-fit test to check whether all the fitted copulas contain no variation. If any copula fails the goodness-of-fit test, adjust the minimum size and repeat the procedure until all the copula are fitted. After several attempts, we divide the data sample into 103 segments with 27 points (trading days) in each segment. There are three segments that fail the goodness-of-fit test at confidence level $\alpha =95\%$, two for Student-t copulas and one for Gaussian copula with test statistics 7.824, 8.097 and 3.973 respectively. However, given that these are weak rejections as we should not reject them at confidence level $\alpha =99\%$ 
and given the risk of losing power if the minimum size is too small, we keep the minimum size as 27 trading days. The merging process is then proceeded layer by layer until all the contiguous copula that come from the same family are merged. Figure \ref{Fig.7} plots the family change of the copula.
From Figure \ref{Fig.7}, we observe that the copula family change among Gaussian, Student-t and Clayton copula several times with the fluctuation of two log-returns of S\&P 500 and Nasdaq indices. It is apparent that the Gaussian copula is the preferable model when the financial markets are stable, while the Student-t copula and Clayton copula appears frequently when the markets are volatile, such as during the period of the financial crisis between 2007 and 2009.

\begin{table}[htbp]
\small
\centering
    \caption{\label{tab:5}Change of copula family and parameter}
      \begin{threeparttable}
        \begin{tabular}{p{3.2cm}<{\centering}p{2.2cm}<{\centering}p{2.7cm}<{\centering}p{2.7cm}<{\centering}p{2.7cm}<{\centering}}
            \hline
            Period & Copula & Parameter & Type of change & Date of change \\
             \hline
             04/01/05-29/04/05 & Gaussian & 0.96 & - & - \\
             
             \multirow{2}*{29/04/05-08/06/05} 
             & \multirow{2}*{Student-t}& 0.93 & \multirow{2}*{Family } & \multirow{2}*{29 Apr. 2005} \\ 
             ~ & & 2.91 &  &  \\ 
             
             08/06/05-17/10/07 & Gaussian & 0.95 & Family & 08 Jun. 2005 \\
             
             \multirow{2}*{17/10/07-22/09/09} 
             & \multirow{2}*{Student-t}& 0.89 & \multirow{2}*{Family } & \multirow{2}*{17 Oct. 2007} \\ 
             ~ & & 2 &  &  \\ 
             22/09/09-07/04/10 & Gaussian & 0.96 & Family & 22 Sep. 2009 \\
             \multirow{2}*{07/04/10-18/10/10} 
             & \multirow{2}*{Student-t}& 0.95 & \multirow{2}*{Family } & \multirow{2}*{07 Apr. 2010} \\ 
             ~ & & 2 &  &  \\ 
             18/10/10-24/11/10 & Clayton & 6.6 & Family & 18 Oct. 2010 \\
             24/11/10-04/01/11 & Gaussian & 0.98 & Family & 24 Nov. 2010 \\
             04/01/11-23/03/11 & Gaussian & 0.95 & Parameter & 04 Jan. 2011 \\
             23/03/11-09/06/11 & Gaussian & 0.97 & Parameter & 23 Mar. 2011 \\
              \multirow{2}*{09/06/11-04/10/11} 
             & \multirow{2}*{Student-t}& 0.92 & \multirow{2}*{Family } & \multirow{2}*{09 Jun. 2011} \\ 
             ~ & & 2 &  &  \\ 
              \multirow{2}*{04/10/11-10/11/11} 
             & \multirow{2}*{Student-t}& 0.89 & \multirow{2}*{Parameter } & \multirow{2}*{04 Oct. 2011} \\ 
             ~ & & 2 &  &  \\ 
              \multirow{2}*{10/11/11-31/01/12} 
             & \multirow{2}*{Student-t}& 0.94 & \multirow{2}*{Parameter } & \multirow{2}*{10 Nov. 2011} \\ 
             ~ & & 2 &  &  \\ 
             31/01/12-31/10/12 & Gaussian & 0.96 & Family & 31 Jan. 2012 \\
              \multirow{2}*{31/10/12-10/12/12} 
             & \multirow{2}*{Student-t}& 0.96 & \multirow{2}*{Family } & \multirow{2}*{31 Oct. 2012} \\ 
             ~ & & 2 &  &  \\ 
             10/12/12-28/02/13 & Gaussian & 0.96 & Family & 10 Dec. 2012 \\
             28/02/13-25/06/13 & Gaussian & 0.97 & Parameter & 28 Feb. 2013 \\
             25/06/13-28/08/14 & Gaussian & 0.95 & Parameter & 25 Jun. 2013 \\
             \multirow{2}*{28/08/14-07/10/14} 
             & \multirow{2}*{Student-t}& 0.96 & \multirow{2}*{Family } & \multirow{2}*{28 Aug. 2014} \\ 
             ~ & & 2 &  &  \\ 
            07/10/14-03/02/15 & Gaussian & 0.95 & Family & 07 Oct. 2014 \\
             03/02/15-01/06/15 & Gaussian & 0.96 & Parameter & 03 Feb. 2015 \\
             \multirow{2}*{01/06/15-02/11/15} 
             & \multirow{2}*{Student-t}& 0.95 & \multirow{2}*{Family } & \multirow{2}*{01 Jun. 2015} \\ 
             ~ & & 2.7 &  &  \\ 
             02/11/15-31/12/15 & Gaussian & 0.97 & Family & 02 Nov. 2015 \\         
             \hline    
        \end{tabular}
         \begin{tablenotes}
        \footnotesize
        \item ``Period'' represents the start and end time of each segment in the form of Day/Month/Year. ``Year'' is denotes by the last two numbers of the year. For Student-t copula, the first parameter is correlation and the second is degree of freedom. ``Family'' and ``Parameter'' represents the change belong to copula family and copula parameter respectively.
      \end{tablenotes}
    \end{threeparttable}
\end{table}
Table \ref{tab:5} displays the detailed information about the change of copulas associated with some financial events:
\begin{itemize}
\item 17 Oct. 2007: Northern Rock, a British bank, faced the severe liquidity risk on 14 September 2007. This is the first signal of the financial crisis and the copula family changes from Gaussian to Student-t copula and lasts almost two years.
\item 7 Apr. 2010: Greek government faced serious financial deficits, and the financial market lost confidence in its solvency on 29 March 2010. One month later, Greek debt was downgraded by S\&P 500 as junk. The copula family changes from Gaussian to Student-t copula.
\item 23 Mar. 2011: Moody's announced to slash the credit rating of the long-term government bond from A1 to A3 on 15 March 2011. After 50 days, ECB decided to bail out the Portugal's market. This leads to the change of copula parameter from 0.95 to 0.97.
\item 31 Oct. 2012: British regulator announced to reform the benchmark rate (LIBOR) and change both the ownership and methodology to avoid the manipulation on 28 September 2012. The copula family changes from Gaussian to Student-t copula. 
\item 01 Jun. 2015: German government bonds were close to collapse between May 2015 and June 2015. This causes the copula family change from Gaussian to Student-t. 
\end{itemize}

These examples verified that the dependence structure of financial asset is likely to change from Gaussian copula to Student-t copula when some negative impact financial incidents occur. Both the copula change and parameter change generally are accompanied by financial events as well.

\subsubsection{Risk measurement strategy}\label{subsec:risk}
\begin{figure}[H]
\centering
\includegraphics[width=12cm]{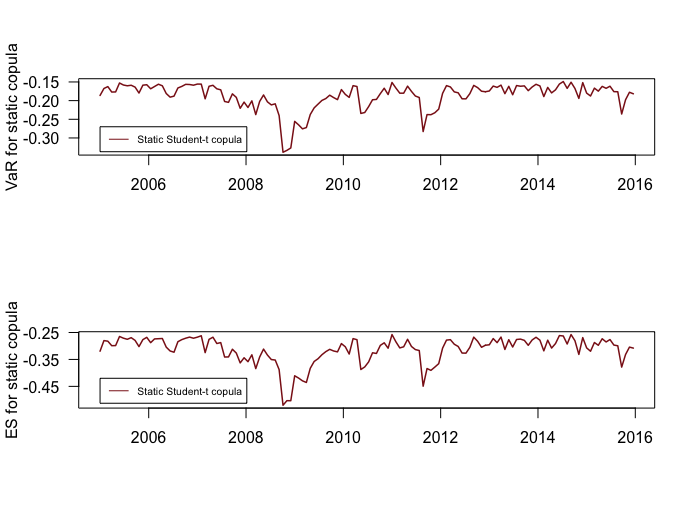}
\caption{VaR and ES based on the static copula for the portfolios of S\&P 500 and Nasdaq indices with equal weight at the confidence level $\alpha =5\%$.}
\label{Fig.8}
\end{figure}

\begin{figure}[H]
\centering
\includegraphics[width=12cm]{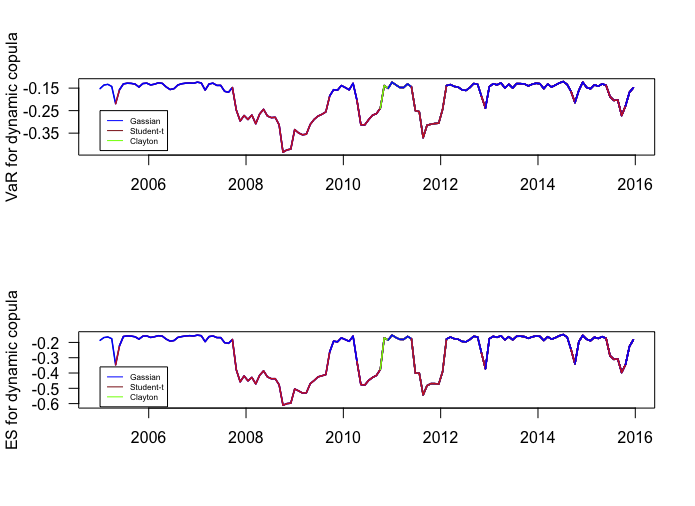}
\caption{VaR and ES based on the dynamic copula for the portfolios of S\&P 500 and Nasdaq indices with equal weight at the confidence level $\alpha =5\%$.}
\label{Fig.9}
\end{figure}

Now, we investigate how the dynamic dependence structure affects risk measures. We assign equal weight to S\&P 500 and Nasdaq indices and then calculate Value-at-risk and Expected shortfall per 20 trading days based on static copula and dynamic respectively to consider the time evolution. Figure \ref{Fig.8} displays the results of the static copula specified in Table \ref{tab:4} that is Student-t (0.94) copula with 2.89 degrees. Figure \ref{Fig.9} presents the outcome that applies the dynamic copula obtained in Table \ref{tab:5} We draw some conclusions through comparison.
\begin{enumerate}
\item Both VaR and ES that calculated by the static copula are more fluctuating than those computed by the dynamic copula. For dynamic copula, it appears distinct fluctuation when the copula family changes from Gaussian to Student-t or Clayton.
\item The change of copula parameter has a significant impact on risk measures. Take the period between 9 June, 2011 and 31 December, 2012 as an example, the maximum VaR and ES estimated by the static copula are 0.29 and 0.42 respectively. However, the dynamic copula forecast 0.35 and 0.55 for VaR and ES, which considers more loss than the static copula. 
\item The absolute value of VaR and ES for the dynamic copula are greater than those for the static copula, especially during 2008 financial crisis, which means that the dynamic copula takes into account more risk. 
\end{enumerate}

\section{Concluding remarks}\label{sec:conclusion}
In this paper, we proposed two new methods to model the dependence structure of financial assets dynamically. Accelerated Moving Window method takes advantage of being applied in real time, which could monitor the change of dependency of financial assets and help to warn the financial crisis. However, this method bears the weakness of delay in detection, which could be improved by adjusting the window size or the warning line in the future. Bottom-up method is proved to be the best the method to detect the change of copula although the appropriate minimum sample size is still a problem. This method is applied to S\&P 500 and Nasdaq indices. The results of risk measures obtained from the dynamic copula illustrate the importance of modeling the dependency of financial assets dynamically.

\end{document}